\documentclass[twoside]{LCWS11}
\usepackage[latin1]{inputenc}
\usepackage[dvips]{graphicx,epsfig,color}
\usepackage{wrapfig,rotating}
\usepackage{amssymb,amsmath,array}
\usepackage{cite}

\newcommand{\met}       {\mbox{$\not\!\!E_T$}}
\newcommand{\mcatnlo}    {\mbox{\textsc{mc@nlo}}}

\begin{document}
\title{
%%%%   Paper title goes here  %%%%%%%%%%%%%%
Top Quark Physics at the Tevatron} %% 
%***********************************************************************
% AUTHORS INFORMATION AREA
%***********************************************************************
\author{Yvonne Peters$^1$
% Optional short acknowledgment: remove next line if non-needed
\thanks{on behalf of the CDF and D0 Collaborations}
% DO NOT MODIFY THE FOLLOWING '\vspace' ARGUMENT
\vspace{.3cm}\\
% Addresses and institutions (remove "1- " in case of a single institution)
1- University of Manchester - School of  Physics and Astronomy \\
Oxford Road, Manchester - UK
%% Remove the next three lines in case of a single institution
}
%%***********************************************************************
% END OF AUTHORS INFORMATION AREA
%***********************************************************************

\maketitle

\begin{abstract}
When the heaviest elementary particle known today, the top
quark, was discovered in 1995 by the CDF and D0 collaborations at the
Fermilab Tevatron collider, a large program to study this particle in
details has started. In this article, an overview of the status of top
quark physics at the Tevatron is presented. In particular, recent
results on top quark production, properties and searches using top
quarks are discussed.
\end{abstract}

\section{Introduction}
With a mass of $m_t=173.2\pm0.6 {\rm (stat)} \pm 0.8 {\rm
  (syst)}$~GeV~\cite{topmassaverage}, the top quark is the heaviest
known elementary particle today. Its most notable properties are the
high mass and its very short lifetime, providing a unique environment
to study a bare quark. The top quark is believed to play a special
role in electroweak symmetry breaking and provide a window to physics
beyond the standard model (SM). 

Since its discovery in 1995~\cite{cdftopdiscovery,d0topdiscovery} by the CDF and D0 Collaborations at the
Fermilab Tevatron collider, a large program to study the top quark in great detail has been initiated at the Tevatron. To
understand whether the observed particle is indeed the top quark as
predicted by theory and to use it for searches of physics beyond the SM
(BSM), it is essential to precisely determine the production
mechanisms and the properties and confront the results with SM
predictions. Deviations of the different quantities from their
prediction could be indications for BSM. Additionally, direct searches
for new physics are performed in the top sector. 

As of today, two particle accelerators provide collisions with enough
energy to produce top quarks: the Tevatron at Fermilab and the Large
Hadron Collider (LHC) at CERN. The Tevatron collider is a
proton-antiproton collider. From 1992 to 1996, Run~I of the Tevatron
was ongoing, providing $p\bar{p}$ collisions at 1.8~TeV
energy. In 2001, Run~II with a collision
energy of 1.96~TeV  started, lasting until September 30th, 2011, and
providing approximately 10.5~fb$^{-1}$ of integrated luminosity for
each experiment. The LHC is a $pp$ collider with currently a center of
mass energy of 7~TeV, that started its operation in 2010. About
5~fb$^{-1}$ of collision data has been provided in 2011. Due to its
high center of mass energy, the production cross section of top quark pairs at
LHC is about a factor 20 higher than at the Tevatron~\cite{atlasdiscovery, cmsdiscovery}.

The large datasets enable us to perform high precision measurements of
top quark production and properties, to study many properties for the
first time and to perform sensitive searches for new physics. 
In this article, an overview of top quark physics at the Tevatron will
be provided. About half of the collected Run~II dataset have been
studied until now.

\section{Top Quark Production}
Top quarks can be produced in pairs via the strong interation or
singly via electroweak interaction. Both interaction modes have been
studied at the Tevatron. In the following recent results of $t\bar{t}$
and single top cross sections will be discussed.

\subsection{Top Quark Pair Production}
At the Tevatron, top quarks are produced to about 85\% via $q\bar{q}$
annihilation and about 15\% through gluon-gluon fusion. The predicted
inclusive $t\bar{t}$ cross section ($\sigma_{t\bar{t}}$) from SM
calculations are of $\sigma_{t\bar{t}}=
6.41\pm0.51$~pb~\cite{ttbarpred1}  and $\sigma_{t\bar{t}}= 7.46 \pm
0.48$~pb~\cite{ttbarpred2} at approximate next to next to leading
order (NNLO) quantum chromodynamics (QCD).

The decay of
the top quark in the SM is to almost 100\% into a $b$-quark and a $W$
boson. We classify different final states of the
$t\bar{t}$ pairs according to the decay of the two $W$-bosons from top
and antitop quark. The main channels we consider for analyses are the
the dilepton final state (5\%),  lepton+jets final state (30\%), and
the all hadronic final state (46\%), where either both $W$-bosons
decay leptonically into an electron or muon, just one decays
leptonically, or none. Channels where at least one $W$-boson decays
into a hadronically decaying tau-lepton are considered separately as
$\tau$+lepton or $\tau$+jets final states, as the identification of
taus is experimentally more challenging. The golden channel at the
Tevatron is the lepton+jets final state, consisting of events with exactly one
isolated electron or muon, at least four jets and large missing
transverse energy,  combines a good ratio
of signal to background with  large statistics and a clear signature.
Events in the dilepton channel have two isolated leptons (electrons or
muon), at least two jets and high missing transverse energy to account
for the two undetected neutrinos. The dilepton final state is very
pure, but suffers from low statistics. Furthermore, the existence of
two neutrinos complicates the reconstruction of the full event
kinematics. In the all-hadronic final state the full event can be
reconstructed, but the channel suffers from high backgrounds due to QCD
mutlijet production. 

The inclusive $t\bar{t}$ cross section has been measured in
lepton+jets, dilepton, all hadronic, $\tau$+lepton, $\tau$+jets and
missing energy plus jets final states. The main tools to
separate $t\bar{t}$ signal from background exploit $b$-jet
identification and the kinematic and topological  differences of
signal compared to background. The $b$-jet identification~\cite{bjettagging} relies
usually on properties of the secondary vertex from $B$-hadron decay or
on tracks displaced with respect to the primary vertex. An example of
both tools, $b$-jet identification and topological information, being used recently is the $t\bar{t}$ cross section
measurement in the lepton+jets final state. At D0 for example,
$\sigma_{t\bar{t}}$ has been measured in lepton+jets using three
different methods~\cite{ljetsd0xsec} using 5.3~fb$^{-1}$ of integrated
luminosity. The first is a counting method using $b$-jet
identification, where events with three jets and at least four jets
are further separated into events with zero, one or at least two
identified $b$-jets. Simultaneously the heavy flavor $k$
factor~\footnote{The heavy flavor $k$ factor defines ratio of the 
   NLO over LO
  $W$+heavy flavor cross sections} of the
dominant $W$+jets background is fitted in order to reduce the
systematic uncertainty. We measure $\sigma_{t\bar{t}}=8.13^{+1.02}_{-0.90}$~(stat+syst)~pb with this
method for a top quark mass of $172.5$~GeV. 
The second method uses no $b$-jet
identification but purely relies on the kinematic and topological
differences of signal and background. A multivariate discriminant is
constructed from several variables showing discrimination between
$t\bar{t}$ signal and $W$+jets background. Using kinematic information
D0 extracts $\sigma_{t\bar{t}}=7.68^{+0.71}_{-0.64}$~(stat+syst)~pb. The third method is a
``combined'' technique where kinematic information is used together
with $b$-jet identification. First, the events are split into events
with two, three or at least four jets and are further devided into
events with  zero, one and at
least two $b$-tagged jets. For events where the background content is
still relatively large a mutlivariate discriminant is formed, which
separates signal from $W$+jets background. Using the combined method
we extract $\sigma_{t\bar{t}}=7.78^{+0.77}_{-0.64}$~(stat+syst)~pb.
All results are limited by systematic uncertainties, and are in good
agreement with theory predictions. The main systematic uncertainties
are the uncertainty from the luminosity calculation, $b$-jet
identification and jet energy scale.  
The CDF Collaboration uses similar methods, and also additionally
employed the method to normalize the measured $t\bar{t}$ cross section
to the cross section of $Z$ production, which is known theoretically
to about 2\% precision. This way, the dominant uncertainty from luminosity
can be reduced. Using 4.6~fb$^{-1}$ of integrated luminosity, CDF
measures $\sigma_{t\bar{t}}= 7.82 \pm 0.38 {\rm (stat)} \pm 0.37 {\rm
  (syst)} \pm 0.15 {\rm (Z~theory)}$~pb for a mass of 172.5~GeV in the
  lepton+jets final state~\cite{ljetscdfxsec}. 

\begin{figure*}[t]
\centering
\includegraphics[height=55mm]{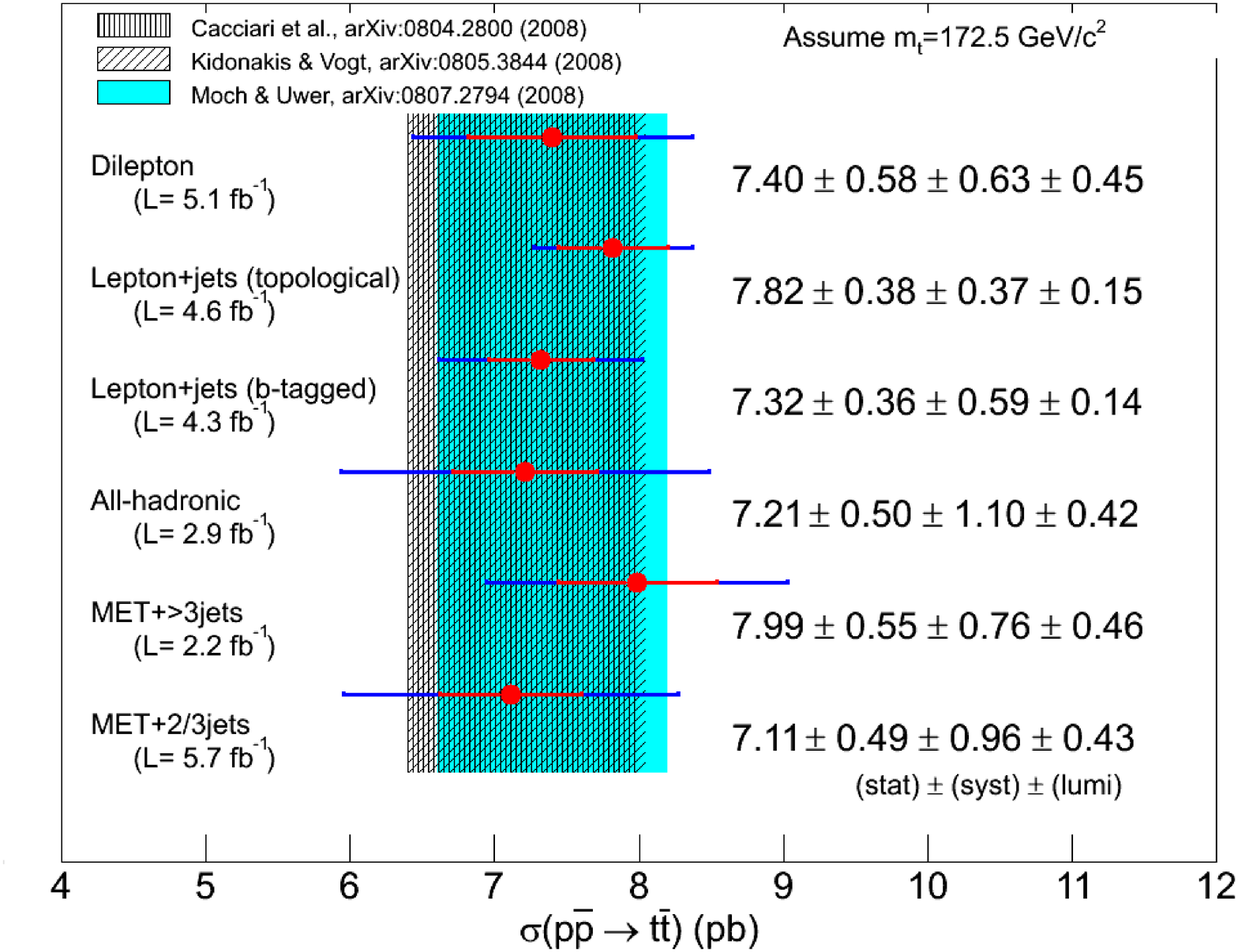}
\includegraphics[height=55mm]{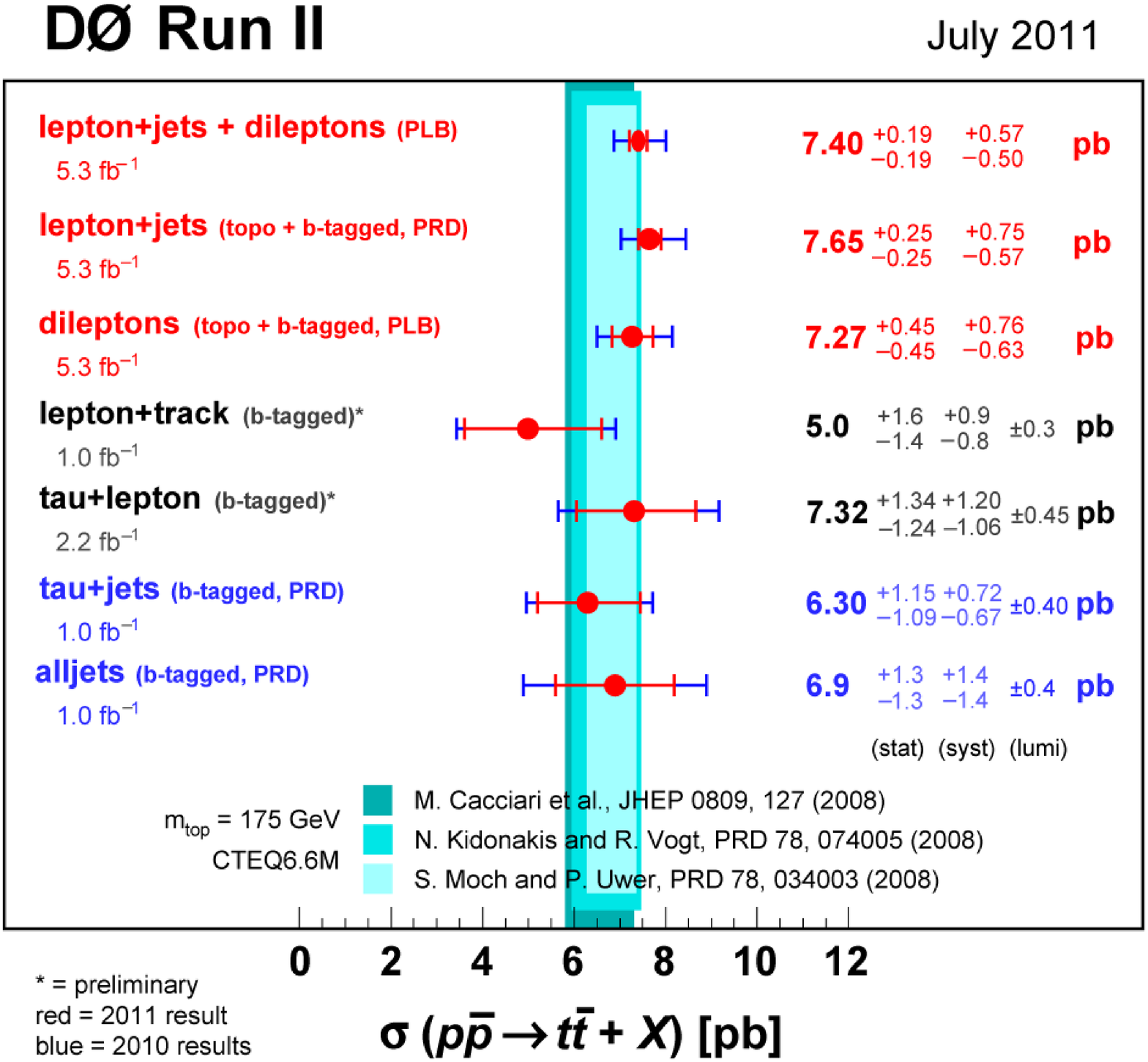}
\caption{$t\bar{t}$ cross sections in various channels by the CDF
  (left) and D0 (right) Collaborations. } \label{xseccombis}
\end{figure*}

For many models of physics beyond the SM, the measured inclusive
$t\bar{t}$ cross section in the different final states could differ
from the theory prediction. Therefore, it is important to extract
$\sigma_{t\bar{t}}$ in the various final states and compare the
results between each other and with theory prediction. Figure~\ref{xseccombis} 
shows the most recent results for $\sigma_{t\bar{t}}$ in various final
states, measured by CDF and D0. All results are in good agreement with
theory predictions as well as between each other.

%mention CDF's Z xsec ratio
%mention all channels
%say that CDF's and our combi with dilepton have uncertainties similar
%to theory 

\subsection{Single Top Quark Production}
Single top quark production happens via the electroweak interaction
and occurs via the $s$-channel, $t$-channel and $Wt$-channel. The
latter has a negligible cross section at the Tevatron. In 2009, single
top quark prodution was observed for the first time by CDF and
D0~\cite{cdfsingletop,d0singletop}, where the $s+t$ channel cross
section ($\sigma_{s+t}$) was measured using up to 3.2~fb$^{-1}$ and
2.3~fb$^{-1}$ of data, respectively. Even though the cross section of the
$s+t$-channel is only about a factor of two smaller than $t\bar{t}$
production, its signature is very similar to $W$+jets events, and
therefore advanced multivariate techniques have to be employed to
distinguish single top signal from background. In particular, boosted
decision trees, neural networks and Bayesian neural networks as well as
matrix element techniques have been used, and the result of the
different methods have been combined. Recently, the D0 collaboration
updated the measurement of the single top $s+t$-channel cross section
using 5.4~fb$^{-1}$ of data, extracting $\sigma_{s+t}=3.43^{+0.73}_{-0.74}$~(stat+syst)~pb~\cite{singletopupdate}.

Since BSM could affect the contributions to $s$-
and $t$-channel differently, it is important to also measure these two
production modes individually. Both collaborations therefore also
perform two dimensional measurements, where the $s$- and $t$-channel
cross sections are measured simultaneously~\cite{cdffullsingletop,d0fullsingletop}.
Recently, D0 reported first observation
of $t$-channel single top production~\cite{tchannelobservation} using 5.4~fb$^{-1}$, obtaining $\sigma_{t}=2.90\pm0.50$~(stat+syst)~pb.

\section{Top Quark Properties}
In order to understand the top quark in detail and to use it for BSM
searches, its properties have to be measured precisely. The large
datasets collected at the Tevatron enable the  measurement of
several properties with high precision, while others can be studies
for the first time. In this section, a selection of recent results are discussed, in particular the top quark mass, the top
antitop mass difference, the helicity of the $W$-boson in top decays,
$t\bar{t}$ spin correlations and the $t\bar{t}$  forward backward
asymmetry. 

\subsection{Top Quark Mass}
The top quark mass, $m_t$, is a free parameter in the SM. Together
with the $W$-boson mass, it sets constraints on the SM Higgs boson. 

With the goal to measure the top quark mass with high precision,
several techniques have been developed. The simplest method is the
template method, where top quark mass dependent templates are
constructed and fitted to the data. In lepton+jets events, the full
event kinematics can be reconstructed by using kinematic fitting
techniques that constrain the invariant mass of the charged lepton and
the neutrino from the leptonic $W$-boson decay to the known $W$-boson
mass. In dileptonic final states, the kinematics are underconstrained
by the two neutrinos, and additional integration over the unknown
quantities is necessary. Several methods exist for this integration,
as for example  matrix weighting or neutrino weighting techniques.   
In the full hadronic final state, the kinematics of the event is fully
known and the main complication arises form the large background and
the large number of possible permutations of jets to match the top
and antitop quarks. 

The second and most precise technique to measure the top quark mass is
the Matrix Element (ME) method. The full kinematic information of each
event is extracted by calculating per-event signal probabilities
$P_{sig}(x;m_t)$ and background probablities $P_{bkg}(x)$, where $x$ are
the momenta of the final state partons. Each probability is calculated
by integration over the leading order (LO) matrix element for $t\bar{t}$
production or background, folded with parton distribution functions
and transfer functions. The transfer functions describe the transition
of the parton momenta as used in the leading order matrix element into
the measured momenta $x$. The top quark mass is obtained by
maximizing the likelihood constructed of a product of the per-event
probabilities. Finally, ensemble tests are performed, as the use
of only leading order matrix element and approximations in the
calculation of the background probabilities requires the method to be
be calibrated. 
A third commonly used method is an approximation of the ME method and
is called ideogram technique. Instead of using matrix elements,
per-event probabilities are calculated using kinematic fitters. 

Additionally to these techniques, a variety of different methods has
been explored at the Tevatron, as for example the extraction of $m_t$
using the transverse momenta of the lepton or using secondary vertex
information. All of these methods are still very limited by
statistical uncertainties, but have the advantage of different
systematic uncertainties being important. These different methods will
be more interesting with much larger datasets. 

The largest systematic uncertainty on the top quark mass using the
three described methods arises from
the jet energy scale (JES). In the lepton+jets and all hadronic final
states, the JES can be fitted in-situ by constraining the invariant
mass of the two jets from the $W$-boson to the known $W$-boson
mass. In dilepton final states, the in-situ JES fit can not be
performed, but recently CDF performed a simultaneous measurement of
$m_t$ in the diletpon and lepton+jets channel, where the fitted JES
from the lepton+jets final state can be applied to the jets in the
dilepton channel~\cite{dilepmasscdf}.

\begin{figure*}[t]
\centering
\includegraphics[height=55mm]{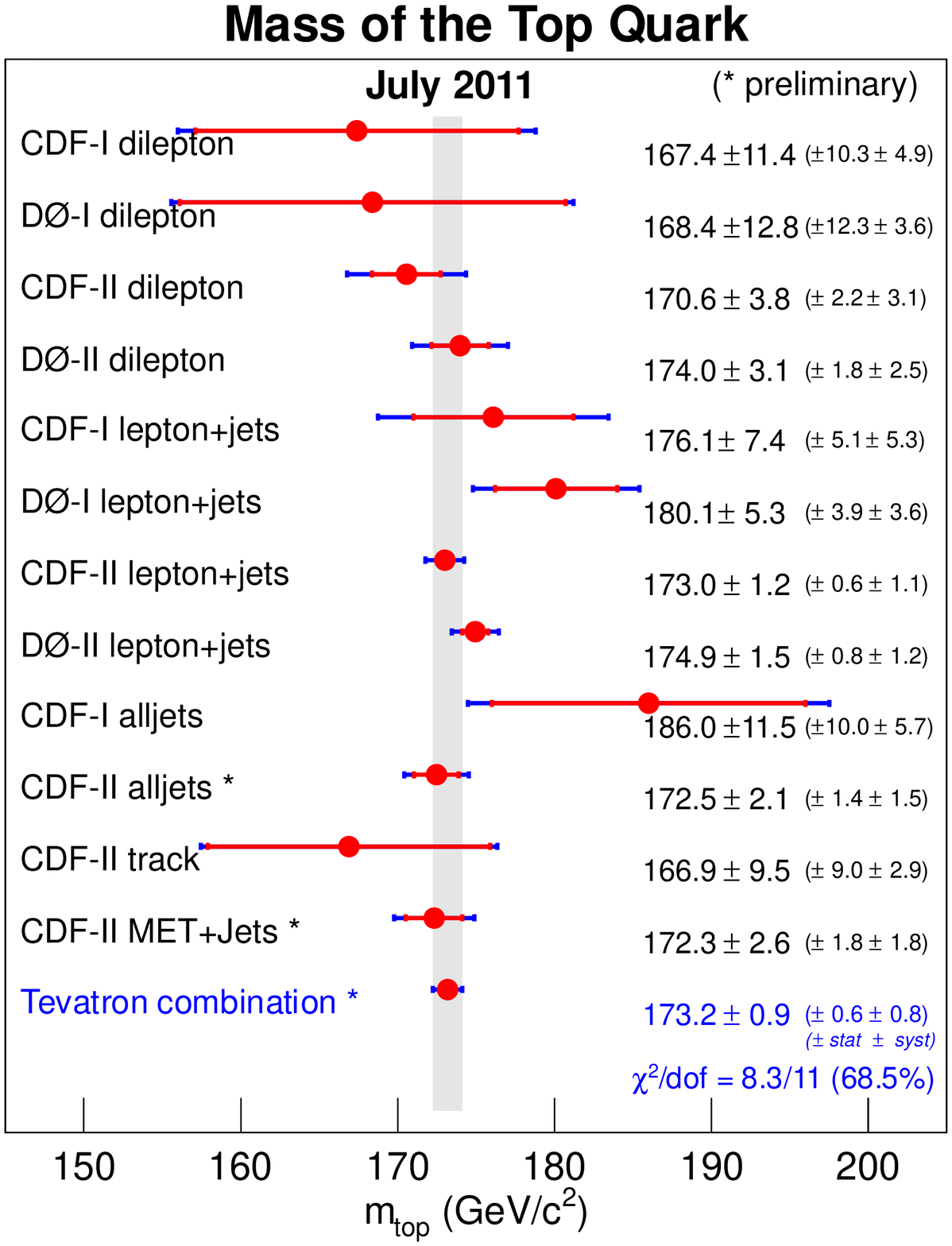}
\includegraphics[height=55mm]{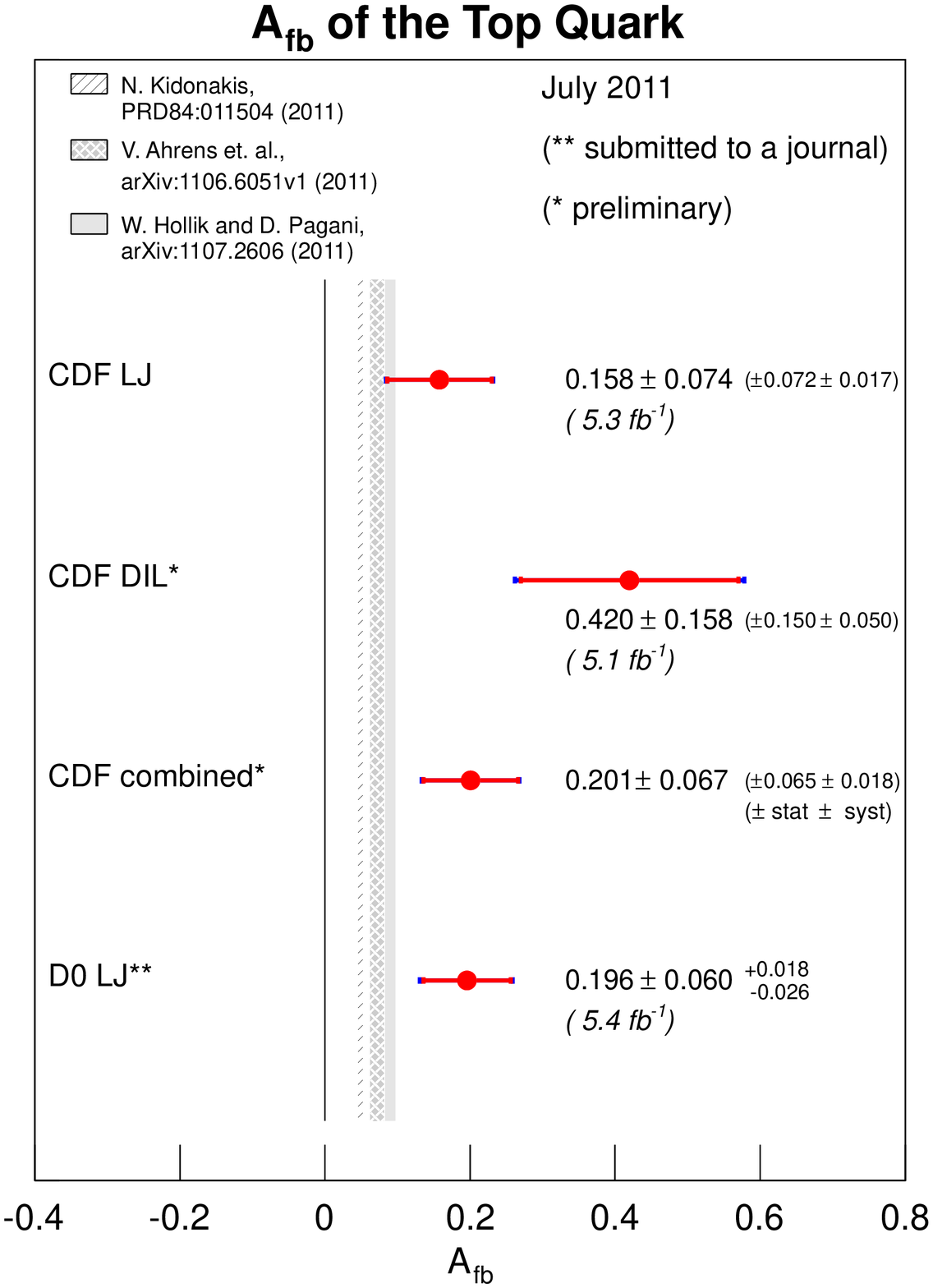}
\caption{Left: Tevatron top quark mass measurements in the different
  final states during Run~I and Run~II and their
  combination~\cite{topmassaverage}. % Right:
  Measurements of the forward backward charge asymmetry $A_{fb}$ at
  the Tevatron~\cite{dilepljetsasymcdf}. 
} \label{masscombiplots}
\end{figure*}

During the life of the Tevatron, the various techniques have been
developed, improved and used to measure the top quark mass with high
precision. Recent measurements of $m_t$ using template techniques are
performed by CDF in the alljets ($m_t=172.5 \pm 2.0 {\rm
  (stat+syst)}$~GeV~\cite{alljetsmasscdf} using 5.8~fb$^{-1}$),
dilepton ($m_t=170.3 \pm 3.7  {\rm (stat+syst)}$~GeV~\cite{dilepmasscdf} using 5.6~fb$^{-1}$) and \met+jets
($m_t=172.3 \pm 2.6  {\rm (stat+syst)}$~GeV~\cite{metjetsmasscdf}
using 5.7~fb$^{-1}$) channels. New results using
the  ME method are a measurement from D0 in the dileptonic final state
($m_t=174.0 \pm 3.0  {\rm (stat+syst)}$~GeV~\cite{dilepmassd0} using 5.4~fb$^{-1}$) and the lepton+jets
final sate ($m_t=174.9 \pm 1.5  {\rm (stat+syst)}$~GeV~\cite{ljetsmassd0}) as well as a measurement in the
lepton+jets channel by CDF ( ($m_t=173.0 \pm 1.2  {\rm
  (stat+syst)}$~GeV~\cite{ljetsmemasscdf} using
5.6~fb$^{-1}$),  the latter being the single most precise measurement
of the top quark mass to date. A combination of all top quark mass
measurements at the Tevatron has been done, resulting in  $m_t=173.18 \pm 0.56 {\rm (stat)}  \pm 0.76 {\rm
  (syst)}$~GeV~\cite{topmassaverage}. The relative precision of 0.6\%
exceeds initial Tevatron expectations. The measured top quark mass is
dominated by systematic uncertainties, where the main sources come
from the differences of the JES for different jet
flavors  and uncertainties on the signal
modeling. The latter include uncertainties on initial and final state radiation, color reconnections, and
next-to-leading order (NLO) versus LO Monte Carlo (MC) generators. In Fig.~\ref{masscombiplots}~(left)  the different Tevatron top quark mass
measurements and the combination are shown.

The direct meausrements of $m_t$ rely heavily on Monte Carlo (MC)
simulations, either for the construction of the templates or the
calibration. Currently used MC simulations are performed in LO QCD, with higher order effects being simulated through parton
showers  at modified leading logarithms level. The top quark mass is a
convention dependent parameter beyond LO QCD. Therefore it is
important to know how the result of direct top quark mass measurements
can be interpreted in terms of renormalization conventions. Currently,
it is still under theoretical investigations how the measured top
quark mass from MC and the top quark pole or ${\overline {MS}}$ mass are
related. The D0 Collaboration has recently performed a determination
of the top quark mass from the measurement of $\sigma_{t\bar{t}}$, by
comparing the measured $t\bar{t}$ cross section to inclusive cross section
calculations versus top quark mass. This allows an 
unambiguous interpretation of the extracted top quark mass  in the pole or ${\overline {MS}}$ mass
scheme~\cite{d0massfromxsec}. Using the pole mass for inclusive cross section calculations
D0 extracted a pole mass of, for example, $m_t=167.5^{+5.2}_{-4.7}$~GeV for the
cross section calculation from Ref.~\cite{ttbarpred2}. Performing the
same extraction, but using a calculation in the ${\overline {MS}}$ mass
scheme yields about 7~GeV smaller values for $m_t$.

\subsection{Top Antitop Mass Difference}
The CPT theorem requires the particles and their antiparticles to have
equal masses. Thus, in direct top quark mass measurements the top and
antitop quark are assumed to be of identical mass. Recently, the D0
and CDF Collaborations have performed measurements of the top antitop
quark mass difference by dropping the assumption of both being of
equal mass, and therefore testing the CPT theorem in the top quark
sector. By extending the event probabilities $P_{sig}(x;m_t)$ to
$P_{sig}(x;m_t, m_{\bar{t}})$, the D0 Collaboration performed the
first measurement of  the mass difference between a bare quark
and its antiquark using the ME method on 1~fb$^{-1}$ of data in the
lepton+jets final state~\cite{firstmassdiffd0}. This 
measurement was updated on 3.6~fb$^{-1}$, yielding $m_t-m_{\bar{t}}=0.8 \pm 1.8 {\rm (stat)} \pm 0.5 {\rm
  (syst)}$~GeV~\cite{massdiffd0}, which is consistent with the SM. The
CDF collaboration performed the mass difference measurement using a
template technique in the lepton+jets channel using 5.6~fb$^{-1}$ of data, resulting in $m_t-m_{\bar{t}}=-3.3 \pm 1.4 {\rm (stat)} \pm 1.0 {\rm
  (syst)}$~GeV~\cite{massdiffcdf}.

\subsection{$W$-Boson Helicity in Top Quark Decays}
In the SM, $W$-bosons couple purely left-handed to fermions, and
therefore constrain the  relative orientation of the spin of the $b$-quark and the $W$
boson from the top quark decay. In NNLO QCD, the fractions
of negative ($f_{-}$), zero ($f_0$) and positive ($f_{+}$) helicity of
the $W$-boson are predicted to be $f_{-}=0.685\pm0.005$,
$f_0=0.311\pm0.005$ and
$f_{+}=0.0017\pm0.0001$~\cite{nnloheli}. Deviations of these values
could indicate new physics contributions. The $W$-boson helicity
fractions have been measured by CDF and D0 in the leptin+jets and
dilepton final states  using template or ME
techniques.  In the template method, the angle $\theta^{*}$ between the
down-type decay product of the $W$-boson and the top quark in the $W$-boson 
rest frame is measured, and the cosine of this angle is fitted to data. To keep the analysis as model-independent as
possible, the fractions $f_0$ and $f_{+}$ are fitted simultaneously,
only constraining the sum of all three fractions to be one.  The CDF
collaboration also uses the ME method to measure the $W$-boson
helicity, where the per-event signal probabilities $P_{sig}$ are
calculated as function of $f_0$ and $f_{+}$. Recently, a combination f
the CDF and D0 measurements has been performed, combining  a D0 measurement
in the dilepton and lepton+jets channel using 5.4~fb$^{-1}$, a CDF
measurement in the lepton+jets final state using 2.7~fb$^{-1}$, and a
CDF analysis in the dilepton final state using
5.1~fb$^{-1}$~\cite{whelitevcombi}. Fitting  $f_0$ and $f_{+}$, the
combination yields $f_0=0.732 \pm 0.063 {\rm (stat)} \pm 0.052 {\rm (syst)}$ and
$f_{+}=-0.039 \pm 0.034 {\rm (stat)} \pm 0.030 {\rm (syst)}$, in good
  agreement with the SM prediction.  Furthermore, the CDF collaboration
  updated the measurement in the dilepton final state using
  5.1~fb$^{-1}$, additionally improving the sensitivity by applying $b$-jet
  identification. This result is not yet included in the Tevatron
  combination and yields  $f_0=0.71^{+0.18}_{-0.17} {\rm (stat)} \pm
  0.06 {\rm (syst)}$ and
$f_{+}=-0.07 \pm 0.09 {\rm (stat)} \pm 0.04 {\rm (syst)}$~\cite{cdfdilepnew}.

\subsection{$t\bar{t}$ Spin Correlations}
While the top quarks are produced unpolarized at
hadron colliders, the spins of the top and antitop quarks are expected
to be correlated. Due to the top quark's short lifetime, the
information of the spin of the top quark is preserved in
its decay products, enabling the measurement of the spin correlation
of the 
top and antitop quark in $t\bar{t}$ events. Recently, $t\bar{t}$ spin
correlations has been measured using a template and ME based  method. 

The template based methods are based on the 
 fact
that the doubly differential cross section, $1/\sigma \times d^2
\sigma /(d \cos \theta_1 d \cos \theta_2)$ can be written as $1/4
\times 
(1-C \cos \theta_1 \cos \theta_2)$, where $C$ is the spin correlation
strength, and $\theta_1$ ($\theta_2$) is the angle of the down-type
fermion from the $W^{+}$  ($W^{-}$) boson or top (antitop) quark decay in the top
(antitop) quark rest frame with respect to a quantization axis. Common
choices are the helicity basis, where the quantization axis is the flight direction of the top
(antitop) quark in the $t\bar{t}$ rest frame,  the beam basis, where the
quantization axis is the beam axis, and the off-diagonal basis, which yields the helicity
axis for ultra-high energy and the beam axis at threshold. The SM prediction for the spin correlation strength $C$
depends on the collision energy and the choice of quantization axis,
and is $C=0.78$ for the Tevatron in the beam basis at NLO~\cite{bernreutherspin}. 
 The spin correlation strength $C$ can be presented as the number
of events where top and antitop have the same spin direction minus the number of events with
opposite spin direction, normalized to the total number of $t\bar{t}$ events, multiplied with a factor
representing the analyzing power of the down-type fermion used to
calculate the angles. The latter factor is one for leptons and
down-type quarks from the $W$-boson decay at LO QCD, and smaller for
up-type quarks and the $b$-quark from  top quark decay. Since it is
experimentally challenging to distinguish up-type from down-type
quarks, the dilepton channel is best to perform the measurement of
$t\bar{t}$ spin correlations. Both,  CDF and D0 Collaborations have
performed a measurement of $C$ by fitting templates for $C=0$ and
the SM value of $C$ of the distribution  $\cos
\theta_1 \cos \theta_2$ to data. Using 2.8~fb$^{-1}$ at CDF and
5.4~fb$^{-1}$ at D0, the measurement of $C$ in the beam basis yields
$C=0.32^{+0.55}_{-0.78} {\rm (stat+syst)}$~\cite{cdfdilepspin} and
$C=0.10 \pm 0.45 {\rm (stat+syst)}$~\cite{d0dilepspin}, in agreement
with SM prediction. Similar to these two analyses in the dilepton final
state, CDF performed the first extraction of $t\bar{t}$ spin
correlations by fitting templates of equal and opposite $t\bar{t}$
helicity  to data. The measured quantity is then translated into
$C$. Using a dataset of 4.3~fb$^{-1}$, CDF measured $C=0.72 \pm 0.64
{\rm (stat)} \pm 0.26 {\rm (syst)}$ in the beam basis~\cite{cdfljetsspin}.

The D0 collaboration also explored the measurement of $t\bar{t}$ spin
correlations using a ME based method.  Per-event signal probabilities
$P_{sig}(H)$ are calculated using matrix elements that include
spin correlations ($H=c$) and do not include spin correlations
($H=u$), 
and are translated into a discriminant
$R=P_{sig}(H=c)/[P_{sig}(H=c)+P_{sig}(H=u)]$~\cite{melnikovschulze}. By
applying this technique to the same D0 dataset of 5.4~fb$^{-1}$ of dilepton events
as for the template based method, a 30\% improved
sensitivity can be obtained, resulting in  $C=0.57 \pm 0.31 {\rm
  (stat+syst)}$~\cite{d0dilepmespin}. Recently, the matrix element-based method
has been extended to the lepton+jets final state using 5.3~fb$^{-1}$
of D0 data, and by combining the
measurement in 
dilepton and  lepton+jets events first evidence for spin correlation
was reported recently, as $C=0.66 \pm 0.23 {\rm (stat+syst)}$~\cite{d0ljetsdilepmespin}.
All Tevatron measurements are in
  agreement with the NLO SM prediction, and all are still limited by
  statistics.

\subsection{$t\bar{t}$ Asymmetry}
At LO QCD, $t\bar{t}$ production is forward
backward symmetric in quark antiquark annihilation processes. At
higher order, interferences between diagrams that are symmetric and
antisymmetric under the exchange of top and antitop cause a preferred direction of the top and antitop quarks and therefore an
asymmetry. In particular, at NLO, the leading contribution arises from the
the interference between tree level
and box diagrams, which yield a positive asymmetry, where the top quark is
preferentially emitted in the direction of the incoming
quark.  A deviation from the
SM prediction could indicate physics beyond the SM. 

At the Tevatron, where the $t\bar{t}$ production is
dominted by the interaction of a valence quark and a valence antiquark
and therefore the (anti)quark
direction almost always coincides with the direction of the incoming
(anti)proton, the measurement of the forward backward charge asymmetry
is conceptionally easy. The asymmetry is defined in terms of the difference
between the rapidity of the top and antitop quarks, $\Delta y$. The
assignment of the final state particles to top and antitop quarks is  determined by applying  kinematic fitting techniques
to the fully reconstructed $t\bar{t}$ events in the lepton+jets  and
dilepton final states. The charge of the lepton(s) is used to
determine which combination of final state objects belongs to the  top
and which to the  antitop quark. The asymmetry is 
defined as $A_{fb}=[N(\Delta y >0) - N(\Delta y<0)]/[N(\Delta y >0) +
N(\Delta y<0)]$, where $N(\Delta y >0)$ and $N(\Delta y <0)$ are the
number of events with rapidity difference larger and smaller zero. Alternatively, the asymmetry can be extracted from the rapidity
 of the lepton(s) only. This has the advantages
that no complete
    reconstruction of the top and antitop quarks and their decays
is required and that the directions of the charged
leptons can be measured with good resolution, while the disadvantage is that
the direction of the lepton is not fully correlated to the top quark
direction, resulting in a loss of sensitivity. In order to compare to theory predictions, the measured $t\bar{t}$
forward backward asymmetries are corrected for acceptance and
resolution effects to obtain the inclusive generated asymmetry. The
correction is done using a $4 \times 4$ matrix-inversion at CDF and with
regularized unfolding at D0. 

Recently, the CDF collaboration measured an
inclusive generated asymmetry of $A_{fb}=0.158\pm 0.074$ using 5.3~fb$^{-1}$ of
data in the lepton+jets channel~\cite{ljetsasymcdf}, and $A_{fb}=0.420 \pm 0.158$ in the
dilepton final state with 5.1~fb$^{-1}$ of
data~\cite{dilepasymcdf}. The combination of these two measurements
results in $A_{fb}=0.201\pm 0.067$~\cite{dilepljetsasymcdf}. The D0 measurement
with 5.4~fb$^{-1}$ of data in the lepton+jets channel yields
$A_{fb}=0.196 \pm 0.060 {\rm (stat)}^{+0.018}_{-0.026} {\rm
  (syst)}$~\cite{ljetsasymd0}. The results are summarized together
with a selection of theory predictions in Fig.~\ref{masscombiplots}~(right). All results are still dominated by
statistical uncertainties. Comparing the measurement to various
theoretical predictions~\cite{asymtheomix} and the prediction of
\mcatnlo~\cite{mcnlo} MC shows about a two sigma
deviation  towards higher
values of the measurements compared to  the prediction. It is not yet clear whether this deviation comes from new
physics contributions or modeling of the SM or anything else, causing
a strong interest in the asymmetry measurements. Various tests to
check the MC modeling have been performed, as for example a test performed  by the
D0 Collaboration to check  the sensitivity to the modeling of
the transverse momentum of the $t\bar{t}$ system, $p_T(t\bar{t})$. This test showed
that the  asymmetry predicted by several MC generators is indeed sensitive to
$p_T(t\bar{t})$, which will require further investigations in the
future.

 Besides the inclusive measurement, it is interesting to investigate the
dependence of the asymmetry on various variables, as for example the
rapidity or the invariant mass of the top antitop quarks,
$m_{t\bar{t}}$. CDF and D0 investigated the $m_{t\bar{t}}$ dependence
by measuring $A_{fb}$ for regions of $m_{t\bar{t}}<450$~GeV and
$m_{t\bar{t}}>450$~GeV. While in D0 data,  no significant dependence
was observed~\cite{ljetsasymd0}, an excess of about three sigma standard deviation
from the \mcatnlo\ prediction was observed by the CDF collaboration for
$m_{t\bar{t}}>450$~GeV~\cite{ljetsasymcdf}.

\section{Searches in the Top Quark Sector}
Besides precision measurements many sensitive direct searches for physics beyond
the SM are performed in the top quark sector. Several models have been
explored in $t\bar{t}$ or single top final states with different methods, as for example classic bump searches or
using multivariate analysis techniques. For example, searches for
$b'$~\cite{bprimecdf}, $t'$~\cite{tprimecdf, tprimed0},
$Z'$~\cite{zprimecdf, zprimed0}, $W'$~\cite{wprimecdf, wprimed0},
charged Higgs bosons~\cite{hpluscdf, hplusd0}, Higgs bosons in
association with a $t\bar{t}$ pair~\cite{tthcdf, tthd0}, flavor
changing neutral currents~\cite{fcnccdf, fcncd0}
and boosted top quarks~\cite{boostedtopcdf} have been performed. Some
of these searches are still the best limits to date. 

\section{Conclusion and Outlook}
Recent measurements of top quark production and properties by the CDF
and D0 Collaborations have been discussed.  About 10.5~fb$^{-1}$ of data have been collected by
    the CDF and D0 collaborations in Run~II of the Tevatron, which
    ended on September 30th, 2011. About half of this dataset has been
    analyzed so far. The Tevatron experiments plan to analyse
the final dataset for those measurement which are
complementary or competitive to  the LHC results, including the top quark mass measurement, the measurement of the
    forward-backward charge asymmetry and $t\bar{t}$ spin
    correlations.

\section{Acknowledgments}

I would like to thank my collaborators from the CDF and D0
collaborations for their help in preparing the presentation and this
article. I also thank the staffs at Fermilab and
collaborating institutions, and acknowledge the support from STFC.

\section{Bibliography}
% ****************************************************************************
% BIBLIOGRAPHY AREA
% ****************************************************************************

\begin{footnotesize}
% IF YOU DO NOT USE BIBTEX, USE THE FOLLOWING SAMPLE SCHEME FOR THE REFERENCES
% ----------------------------------------------------------------------------

% ----------------------------------------------------------------------------

\end{footnotesize}

\end{document}